# Microstructural and Optical Emission Properties of Diamond Multiply Twinned Particles

Vadim Lebedev,[1,a)] Taro Yoshikawa,[1] Christoph Schreyvogel,[1] Lutz Kirste,[1] Jürgen Weippert,[1] Michael Kunzer,[1] Andreas Graff,[2] and Oliver Ambacher[1]

[1] *Fraunhofer Institute for Applied Solid State Physics, IAF, 79108 Freiburg, Germany*

[2] *Fraunhofer Institute for Microstructure of Materials and Systems, IMWS, 06120 Halle, Germany*

[a)]Author to whom correspondence should be addressed: vadim.lebedev@iaf.fraunhofer.de

## ABSTRACT

Multiply twinned particles (MTPs) are fascinating crystallographic entities with a number of controllable properties originating from their symmetry and cyclic structure. In the focus of our studies are diamond MTPs hosting optically active defects –objects demonstrating a high application potential for emerging optoelectronic and quantum devices. In this work, we discuss the growth mechanisms along with the microstructural and optical properties of the MTPs aggregating high-density of "silicon-vacancy" complexes on the specific crystal irregularities. It is demonstrated that the silicon impurities incite a rapid growth of MTPs via intensive formation of penetration twins on {100} facets of regular octahedral grains. We also show that the zero-phonon-line emission from the Si color centers embedded in the twin boundaries dominates in photo- and electroluminescence spectra of the MTP-based light-emitting devices defining their steady-state optical properties.

## I. Introduction

Synthetic diamond becomes a key material for a number of emerging devices including integrated photonic circuits [1], high frequency electro-acoustic filters [2], power transistors [3], and quantum-effect sensors [4]. Despite the fact that diamond heteroepitaxial layers have been intensively studied since the 1950s using various modifications of chemical vapor deposition (CVD) technique [5], a direct heteroepitaxy is proven challenging. First, this is due to the lack of thermally and lattice-matched substrates. Second, it is caused by a metastable character of the diamond face centered cubic, *fcc*, crystal structure (*Fd3m*). Only the recent advances in a microwave-plasma CVD on iridium substrates via bias-enhanced nucleation [6] (BEN) pave the way for a rapid progress of all-diamond functional heterostructures and nanodevices.

Over the last decade, a particular attention of the researchers focuses on "impurity-vacancy" color centers in diamond, i.e. negatively charged nitrogen- (NV) and silicon-vacancy (SiV) complexes [7]. Embedded into the diamond hetero- and nanostructures, such complexes exhibit attractive properties for quantum technologies [4]. The best-investigated NV centers already demonstrated a nanoscale spatial resolution in ultra-high sensitivity magnetometery [8] and thermometry [9] sensors via optically detected electron spin resonance. Another prospective application of the color centers are atomic-scale light sources offering a number of key advantages for emerging quantum computing and sensor devices, i.e. single photon emission for quantum transducers [7], resolution beyond the diffraction limit for near-field optical imaging [10], high-repetition-rate optical signal sources for optical circuits [11], etc. For such integrated nanoemitters, dimensional and material properties of the host diamond structure ("container") play a key role, greatly affecting photostability, linewidth of the zero phonon line (ZPL), and excited state lifetime along with dephasing and spectral diffusion effects.

In the most published works, the container hosting the nanoemitters has often the shape of a sharp tip fabricated from a bulk crystal by conventional micromachining methods. The color centers are created in the diamond lattice by implantation of the dopant atoms followed by "activation" via a high temperature annealing. The disadvantages of this "top-down" approach are lattice damages caused by the implantation along with a high surface roughness and inevitable deviations from the desired geometry of container due to the dry-etch processing. It leads to unwanted absorption, scattering, and deflection of the emitted photons [10]. Alternatively, *as grown* diamond surfaces and self-assembled nanostructures offer chemically pure and crystallographically well-defined {100} and {111} low-index facets [12]. Hence, the photon emission from these facets can be extracted nearly lossless and detected by photo-luminescence (PL) and electro-luminescence (EL) methods. Such containers provide many benefits for fabrication reliability and operational control of the nanoemitters. However, there are also drawbacks for the "bottom-up" approach such as density fluctuations of the impurities (i.e. caused by segregation) along with size/shape variations of the containers characteristic for continuous films and nanostructures, respectively. These issues originate from the crystal growth and should be moderated to ensure a rapid implementation of the "self-assembled" structures as microscopic light sources.

In our previous work [13], we demonstrated that micrometer-scale *multiply twinned particles* (MTPs) such as diamond decahedron (Dh) and icosahedron (Ih) structures can be reproducibly formed on iridium (100) by means of BEN/CVD techniques in high supersaturation conditions. In this paper, we present the experimental studies on Si-doped diamond MTPs, which provide as a well-defined host crystal structure as well as advantageous nucleation sites for the formation of optically active SiV complexes. We discuss MTP crystallization mechanisms together with the specific PL and EL spectral features obtained from isolated MTPs and light-emitting Schottky diodes fabricated on continuous MTP layers. Particular attention is paid to the role of Si as an anti-surfactant agent, which greatly affects electro-optical and micro-structural properties of the diamond MTPs.

## II. Experimental details

Isolated diamond MTPs and closed MTP layers are deposited on in situ carbonized 2" Si(111) substrates (SiMat, $\rho \approx 20$ mΩ cm) in a modified SDS5200S microwave plasma CVD reactor (Cornes Technologies) operating at a power of 1.4 kW by 2.4 GHz. The surface temperature is measured by a Williamson dual-wavelength infrared pyrometer. Technical details on the CVD equipment along with a general description of the BEN/growth procedures can be found elsewhere [14]. Table I summarizes the experimental conditions for the carbonization, BEN, and CVD processes.

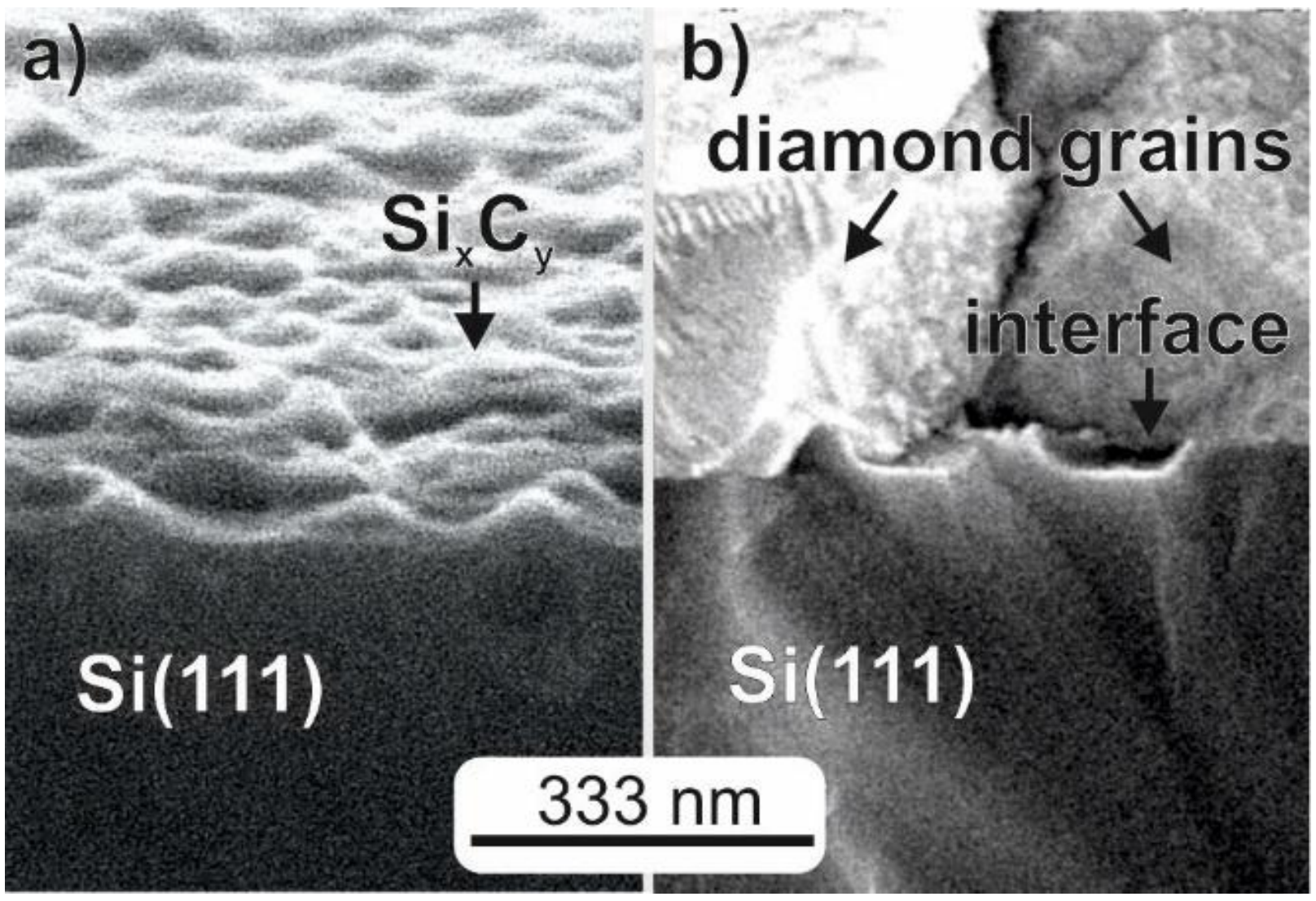

**FIG. 1.** Scanning electron microscopy (SEM) image (75° view angle) of a) "rough" carbonized (111) Si surface and b) hetero-interface between diamond grains and silicon substrate taken after the cleavage along a <110> direction at the central area of the wafer.

Prior to the BEN processing, the Si substrates are carbonized by annealing in $H_2$+$CH_4$ plasma at 900 °C to form an ultra-thin 3C-$Si_xC_y$ layer employed as a buffer for the diamond deposition. The thickness of carbide layer does not exceed 2-3 nm, since carbon bulk diffusion is a self-limited process [15]. It is important to note that the parameters chosen for BEN and CVD processes are kept identical to those used for the MTP growth on iridium (see Ref. [13]) in order to perform a direct comparison of the MTPs growth dynamics and their physical properties. The growth time varies from 60 to 240 minutes in order to obtain MTPs in different growth stages for scanning (SEM) and transmission electron microscopy (TEM) observations as well as for X-ray diffraction (XRD) and secondary ion mass spectrometry (SIMS) analyses. The closed MTP layers for cross-sectional SEM studies are prepared by a cleavage along a <110> direction of the Si substrate.

A scanning confocal setup based on a Renishaw *inVia* Raman microscope is employed for micro-PL (µPL) and micro-EL (µEL) spectra acquisition and light intensity mapping. The µPL measurements of isolated MTPs are carried out at excitation laser intensity of <0.1 µW at 532 nm wavelength. For the µEL studies, edge-type SiV light emitting devices (SiV-LEDs) are fabricated on closed MTP layers by defining circular Au(400 nm)/Ti(5 nm) Schottky contacts using sputter deposition. Diameter of the contacts and thickness of the MTP layer are 500 µm and 2.5 µm, respectively. The manufactured SiV-LED arrays are mounted on the printed circuit board and then wire-bonded to the external pads.

## III. Experimental Results

### A. *In situ* carbonized silicon surface

After loading an "epi-ready" Si(111) substrate into the CVD chamber, a 30 minutes carbonization process is carried out at 900 °C under low pressure and low microwave power conditions (see Table I). The purpose of this treatment is multifold. First, the carbonization produces (111)-oriented 3C-$Si_xC_y$ ultra-thin non-continuous films. Second, it increases the surface roughness of the substrate (Fig. 1a).

As shown elsewhere, the three-dimensional (3D) morphology of the substrate surface might stabilize metastable crystal polytypes by 3D lattice replication [15]. Therefore, the carbonization ensures nucleation and growth of (111)-oriented diamond nuclei. In addition, the carbide film plays a role of protective coating preventing an excessive sputtering of silicon surface by the hydrogen plasma. As shown in Fig. 1(b), the $Si_xC_y$ surface is stable in $H_2$-plasma up to 900 °C retaining its morphological features after numerous diamond deposition circles.

### *B. MTP growth dynamics*

The systematic SEM observations at different stages of the grain growth allows us to distinguish between five main crystallization phases (titled from A to E in Fig. 2a and Table II) occurring before the coalescence stage, and transformation of the isolated diamond MTPs into the continuous MTP layer (Fig. 2b). It is important to note that in contrast to the grains crystallized on Ir at the identical growth conditions, nearly all the (111)-oriented octahedron (Oh) pyramids on carbonized Si(111) substrates are transformed into completed Ih-MTPs via the same twinning mechanism.

As revealed by SEM observations, the Oh(111) grains grow twin-free till the area of the {100} facet surpasses a critical value of $A_c \approx 300\times300$ nm$^2$ (phase A). In other words, for the (111) diamond grains on Si substrates at the actual growth conditions, a specific mean free path of adsorbed carbon species [16,17] involved in twin growth does not exceed the characteristic length $<\iota> \approx 150$ nm. This indicates a strongly kinetic-limited regime of the twin nucleation [13].

It is important to note that the obtained 3D morphology of the carbonized surface supports the growth of (111)-oriented Oh grains, but does not provide favorable conditions for a 3-axial epitaxial relationship. As a result, the BEN-assisted deposition on 3C-SiC/Si(111) produces "fiber-textured" (111) oriented grains, with no dedicated in-plane orientation. This structural irregularity does not affect the properties of the isolated MTPs, but has a radical impact on the closed MTP layer. As shown in Fig. 2b, the surface of the closed thin film consists of two components: triangular fiber-textured Oh(111)-cores and an area occupied by the twinned segments of the MTPs. Such morphology cannot provide the conditions for a further (111)-texture growth and the MTP layer gradually becomes pure polycrystalline.

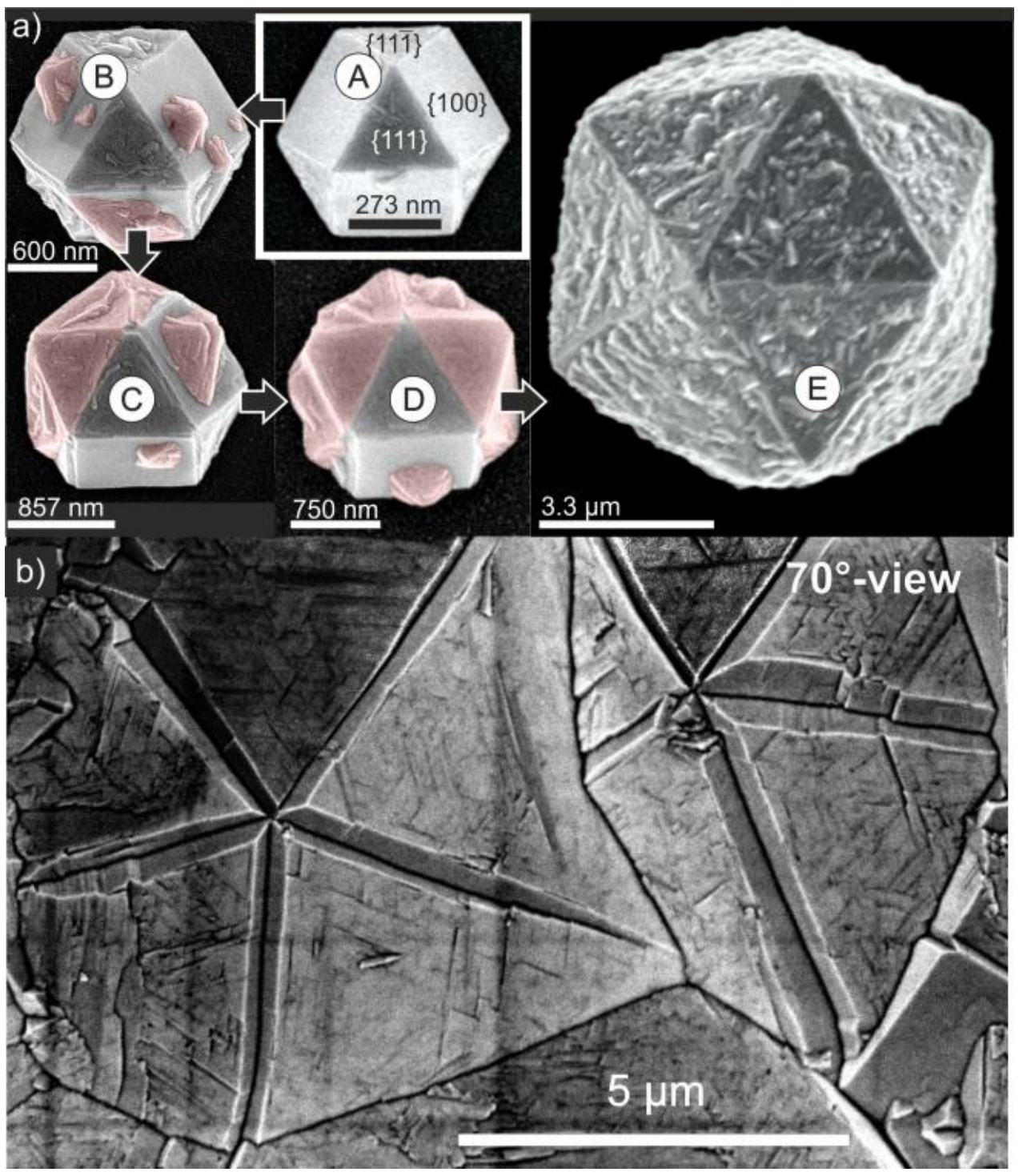

**FIG. 2.** Top-view SEM micrographs demonstrating a) step-by-step transformation of octahedral Oh(111) pyramid (phase A) via "penetration twin" nucleation on {001} facets (phases B-C) into Dh-MTP (phase D) and then into Ih-MTP (phase E). b) SEM image (70° view angle) of a 2.5 µm thick, fiber textured diamond MTP layer at the stage of completed coalescence.

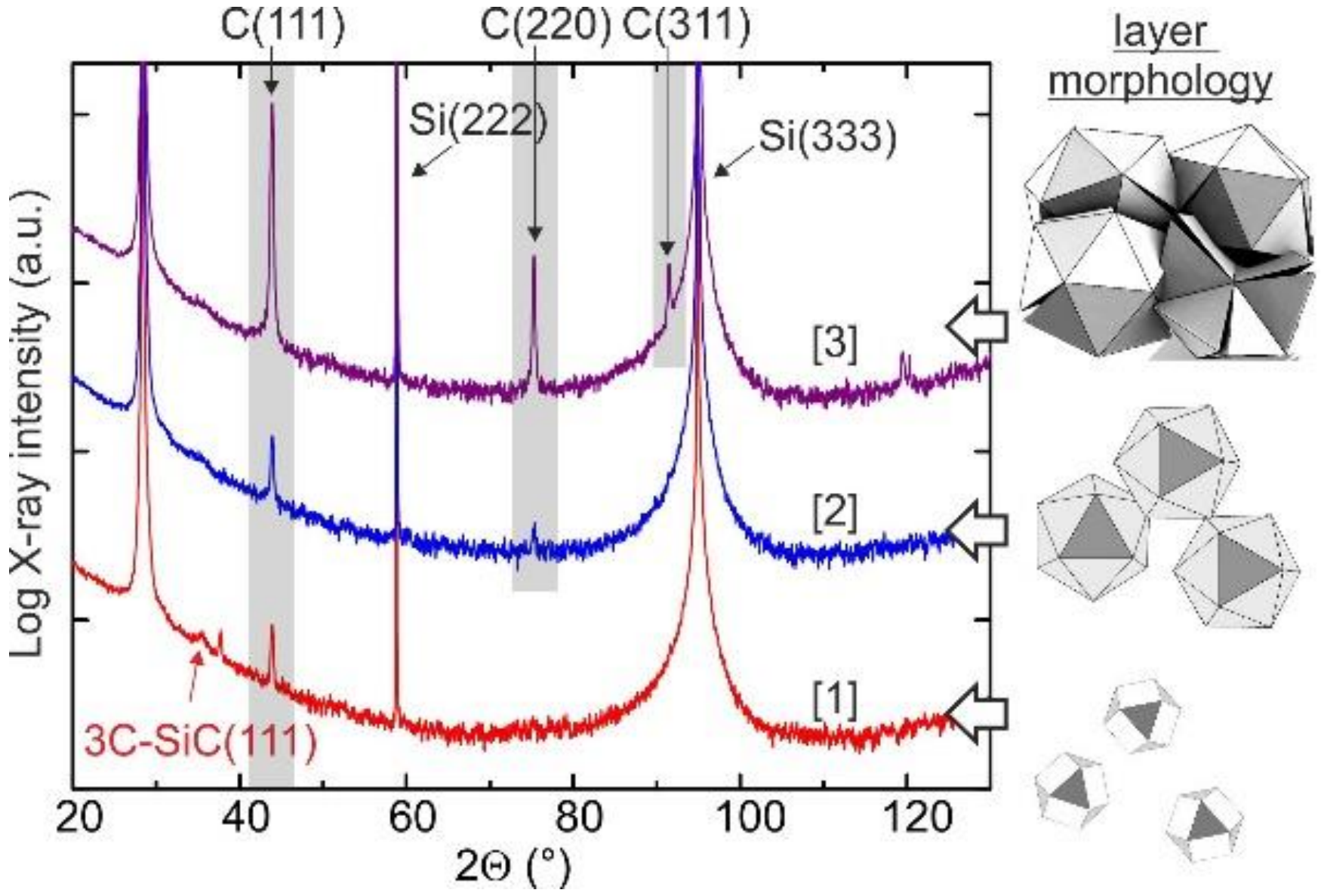

**FIG. 3.** XRD 2Θ/Θ scans of diamond/$Si_xC_y$/Si(111) samples containing: [1] - isolated (111)-oriented Oh and Ih grains (Fig. 2a); [2] - Ih-grains at the stage of partly completed coalescence (Fig. 2b); and [3] – thick diamond film showing polycrystalline structure. Insets: schematic representation of the layer morphologies corresponding to the plotted curves [1]-[3].

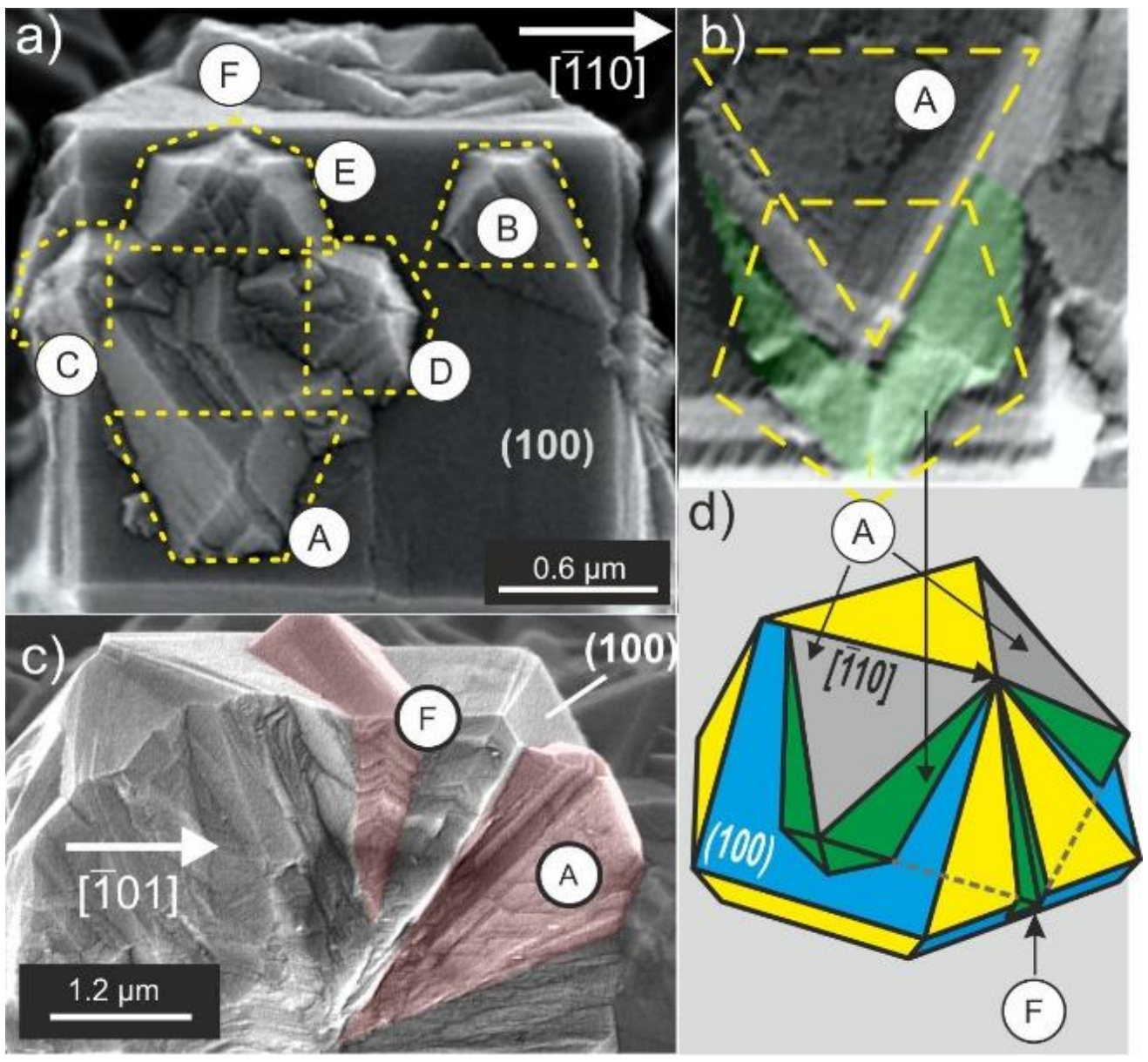

**FIG. 4.** SEM micrographs (70° - view angle) illustrating penetration twin (PT) nucleation and development. a) Initial formation of four orthogonally oriented PTs on a {100} facet of the Oh(111) grain: PT types A and B with a regular 4-fold symmetry, and PT types C, D, E developing a 5-fold (111) twinned structure on the reentrant side-walls. b) Close-up view of a partly developed 5-fold twinned structure. c) Cross-section view of the diamond grain demonstrating two types of PTs: regular one of type A and a twin of type F - the special case of PT penetrating the {111} facet. d) Schematic representation of Ih-related twins with respect to the parent Oh(111) geometry.

The XRD-2Θ/Θ curves plotted in Fig. 3 clarify the SEM observations of the growth phases shown in Fig. 2. Curve [1] represents the insulated island growth resulting in a well-defined (111) fiber-textured structure, with the major impact acquired from the Oh(111) cores. One can see that the layer is still not closed exposing (111) 3C-$Si_xC_y$ surface (a weak reflex at 2Θ ≈ 35°). The coalescence stage stimulates an irregular crystallization on concave (reentrant) surfaces formed by non-coincident (111) facets of adjacent Ih grains (curve [2]). It is manifested by appearance of a 220 diamond lattice reflection. Thicker layers exhibit a polycrystalline morphology (curve [3]).

### C. MTP formation via "penetration" twin mechanism

Fig. 2a demonstrates that an MTP formation is instigated on a (100) plane family of the regular Oh pyramids by nucleation and evolution of the specific crystal irregularities – "penetration" twins (PT). PTs often appear on nominal {100} facets during growth and remain one of the most studied twin types in synthetic diamond [18,19,20]. According to the model suggested by Butler et al. [21], twinned islands might nucleate at local surface regions exposing {111} structures like etch pits on nominal {100} facets (around dislocations) or macrostep edges (on vicinal {100} planes). In earlier works, it was also suggested that PTs might originate from adsorbed impurities on nominal {100} facets (see section IV.B for discussion). Once formed, the PTs continue their growth developing the core of characteristic growth hillocks.

In Fig. 4a, the {100} facet of a regular Oh(111) pyramid is displayed illustrating the PT nucleation and growth processes at the phase B. The example shown here is an extremal case demonstrating various growth phenomena occurring at an initial twinning stage. First, Fig. 4a shows that a simultaneous nucleation of four orthogonally oriented PTs might occur on a nominal {100} facet. This phenomenon is due to a partial compatibility of the dangling bonds of (100) surface and a bottom plane of the PT (see the discussion section). As confirmed by SEM observations (see Fig. 2a), a dominant PT orientation is always compatible with the developing Ih structure (type A in Fig. 4a), while "non-Ih"-oriented PTs (if nucleated) are rapidly overgrown by the multi-twinned tetragonal units forming Dh- (phase D) and then Ih-MTPs (phase E).

Fig. 4c illustrates a penetration character of such twins using a (110) cross-sectional SEM micrograph of the Oh(111) pyramid containing two kinds of PTs: a conventional twin penetrating the {100} facet (type B), and a specific Ih-related PT on the {111} Oh facet (type F) consisting of two (111)-twinned sections. As shown schematically in Fig. 4d, both types of 5-fold-twinned segments are crystallographically related forming cyclic Ih structure after the fusion at the later stages of crystallization.

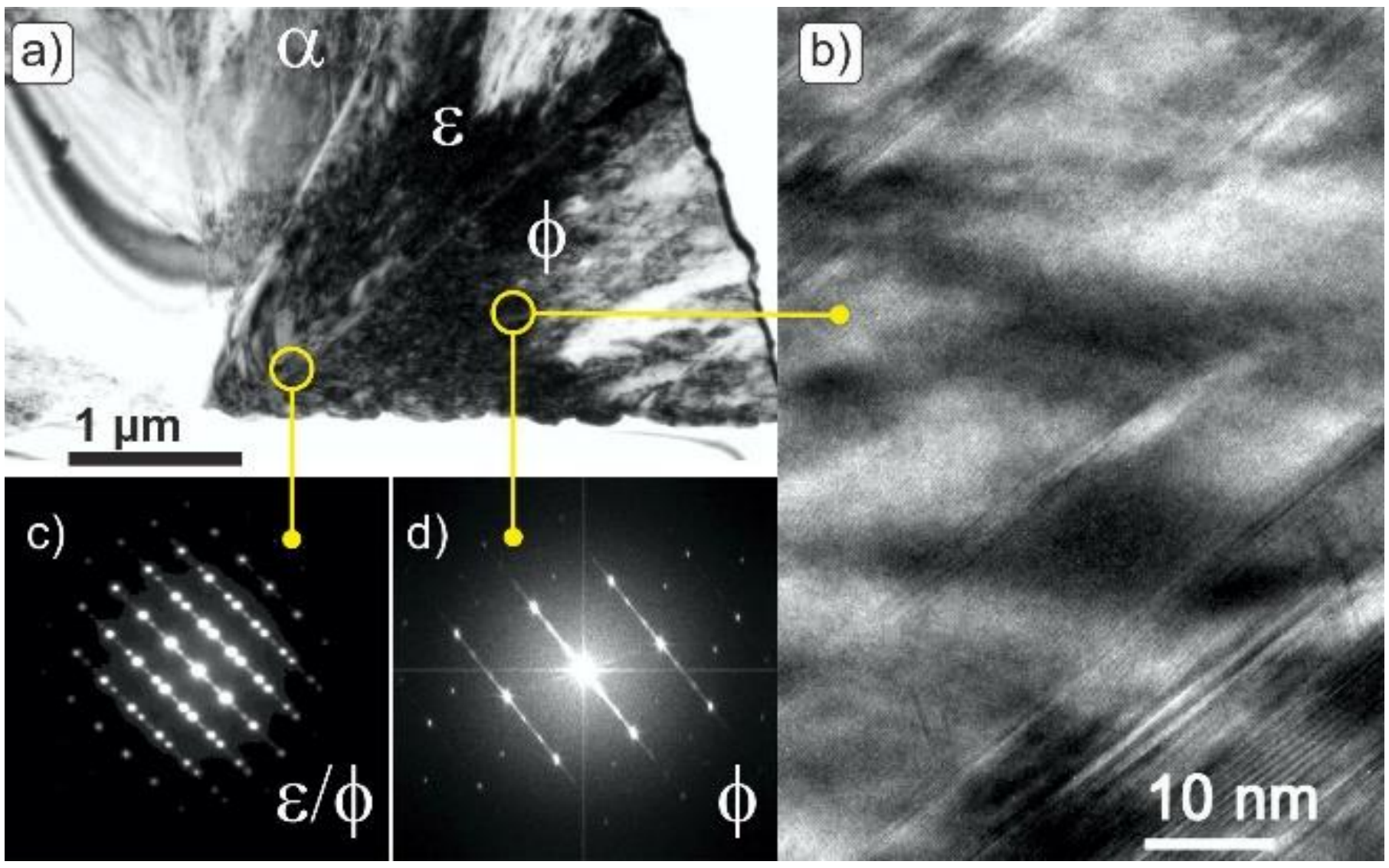

**FIG. 5.** Transmission electron microscopy studies of Ih-MTP: a) micrograph of the Ih-MTP containing epitaxially oriented core ($\alpha$), and two adjacent Ih-twins ($\varepsilon$ and $\phi$); b) HR-TEM image of $\phi$-twin having a high-density of stacking faults (SFs); c)-d) selective area diffraction patterns (SAED) taken c) at twin boundary area and d) at middle area of $\phi$-twin.

Also, as shown in Fig. 4b, the formation of 5-fold structures, which will aggregate and transform into the MTPs at the final stage, starts on the reentrant surfaces formed by crossing {100}´ and {111}´ facets of the PTs and the {100} facet of Oh(111). Similar to the case of (111)-plane twinning on Ir substrates (see Ref. [13]), such concave features support a rapid growth of (111) twins (also called "contact" twins). As a result, PT hillocks on {100} and {111} facets of regular Oh(111) pyramids transform into 5-fold-twinned tetragonal segments forming building blocks for Dh- and Ih-MTPs.

The MTP formation initiated by the PTs results in the specific cyclic structure of the grain. It is characteristic by pronounced (111) twin boundaries (TBs) and by high densities of dislocations (i.e. of stacking faults, SFs) penetrating the twinned sections. Figs. 5a-b show the results of TEM analysis carried out on an isolated Ih-structure consisting of the epitaxially oriented core ($\alpha$) and two attached Ih-twins, $\varepsilon$ and $\phi$. Here, the SAED pattern taken at the $\varepsilon/\phi$ TB area (Fig. 5c) confirm an angular relationship between the adjacent twins, and another one taken at the middle area of $\phi$-domain (Fig. 5d) indicate a high density of (111) SFs specific for the twinned fcc crystals [7]. HR-TEM studies reveal the dense SF network from the same twin area (Fig. 5b). Meanwhile, a limited sensitivity of TEM-based energy-dispersive X-ray spectroscopy ($\approx$0.1 at.%) does not allow silicon detection within the investigated MTPs.

### D. Formation and properties of SiV centers

Exposed to the $H_2$-plasma a silicon surface is a natural source of background Si doping for the diamond MTPs [12]. We can consider two dominant mechanisms of Si atomic transport into the MTP domain: i) diffusion from the substrate via dislocations, grain and twin boundaries [22], and ii) re-deposition of silicon atoms sputtered from the substrate surface exposed to the plasma. The contributions of these mechanisms to the Si concentration depends on the MTP layer thickness. As revealed by the SIMS depth-profiles measured on closed MTP layers (Fig. 6ab), the Si concentration decays nearly exponentially with a rising film thickness. We can assume that for non-continuous layers and isolated MTPs, the Si re-deposition is a dominant doping mechanism, while in the closed MTP layers the out-diffusion via extended defects governs the silicon doping.

Despite of a limited spatial resolution, the SIMS mapping highlights the grain boundary area as the main diffusion path and the segregation location for the Si atoms (Fig. 6c, position A), while the Si concentration within the MTP is one order of magnitude lower comparing to the boundary region. In order to improve the analysis accuracy, we used μPL and μEL mapping, which has a substantially higher XYZ-resolution and a lower detection limit.

### E. Stimulated light emission from MTP-SiV centers

Fig. 7 summarizes the results of μPL studies on diamond grains crystallized on SiC/Si(111) surface. As shown in an overview map (Fig. 7a), the isolated MTPs are the most intensive sources of SiV "zero-phonon-line" (ZPL) emission centered at $\lambda\approx$738 nm (300 K). A typical PL spectrum taken from the MTP region at 300 K reveals a full width at half maximum (FWHM) of $\approx$4.8 nm for the SiV ZPL (Fig. 7b). PL spectrum measured at 10 K shows a ZPL's "blue shift" and FWHM of 2.8 nm (the observed linewidth is limited by the resolution of the spectrometer). This is an expected behavior, which was methodically discussed in many recent works on low-temperature behavior of SiV complexes (e.g. Ref. [23]).

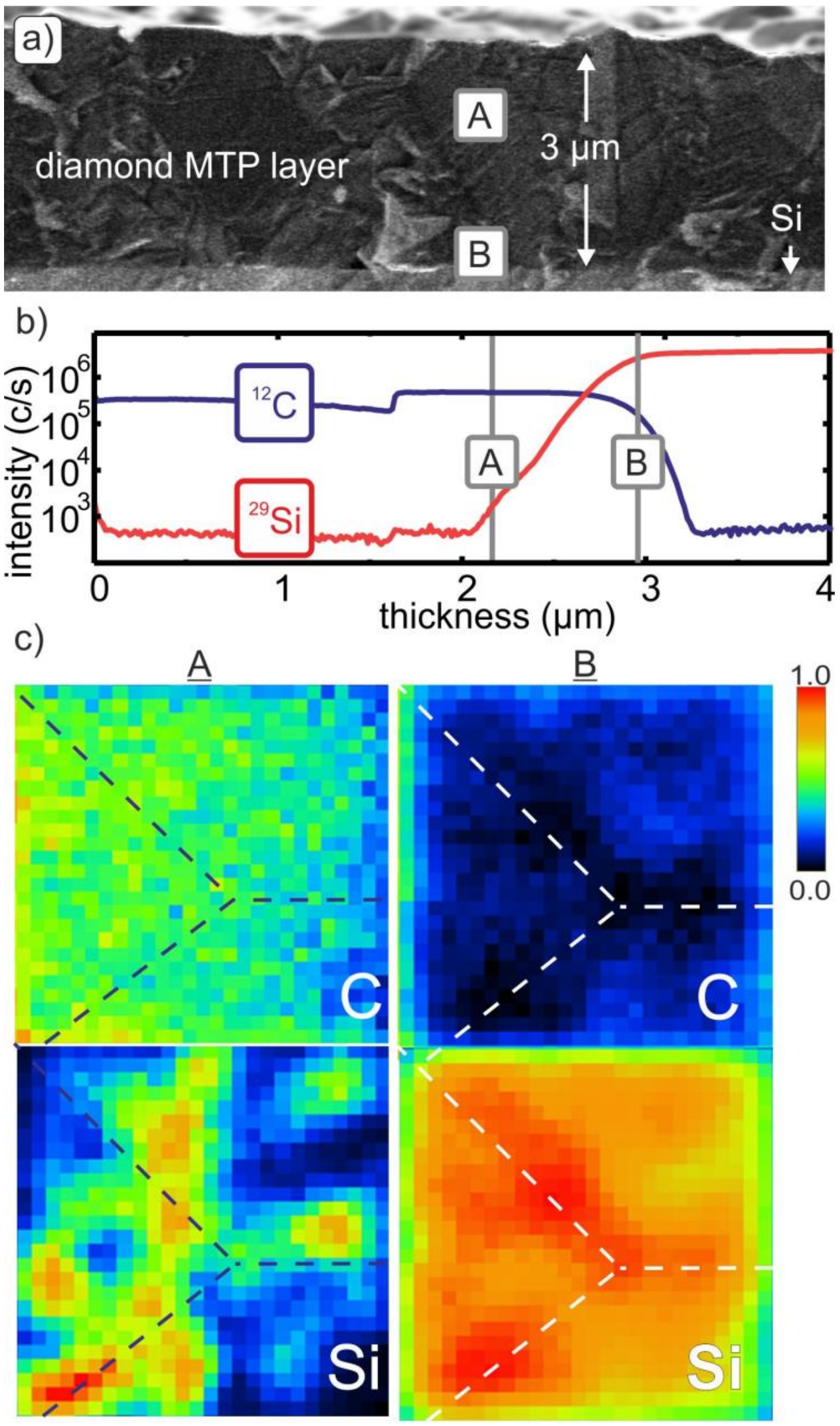

**FIG. 6.** a) SEM micrograph (70° - view angle, (110)-cleavage) of the investigated diamond MTP-film. b)-c) SIMS measurements carried out on closed MTP layers. b) SIMS depth-profiles for $^{12}C$ and $^{29}Si$ masses. c) Normalized in-plane mass distribution maps ($50\times50$ µm$^2$): position A – measurement point at a distance of $\approx$1 µm from the heterointerface, position B – diamond/Si heterointerface area; the dashed lines are a guide for the eye indicating approximate positions of the grain boundaries.

The most interesting finding is that the outer MTP grain boundaries, which have a significantly higher Si concentration according to the SIMS maps (Fig. 6), hardly emit any light at around 738 nm. In contrast, the SiV centers emitting intensive

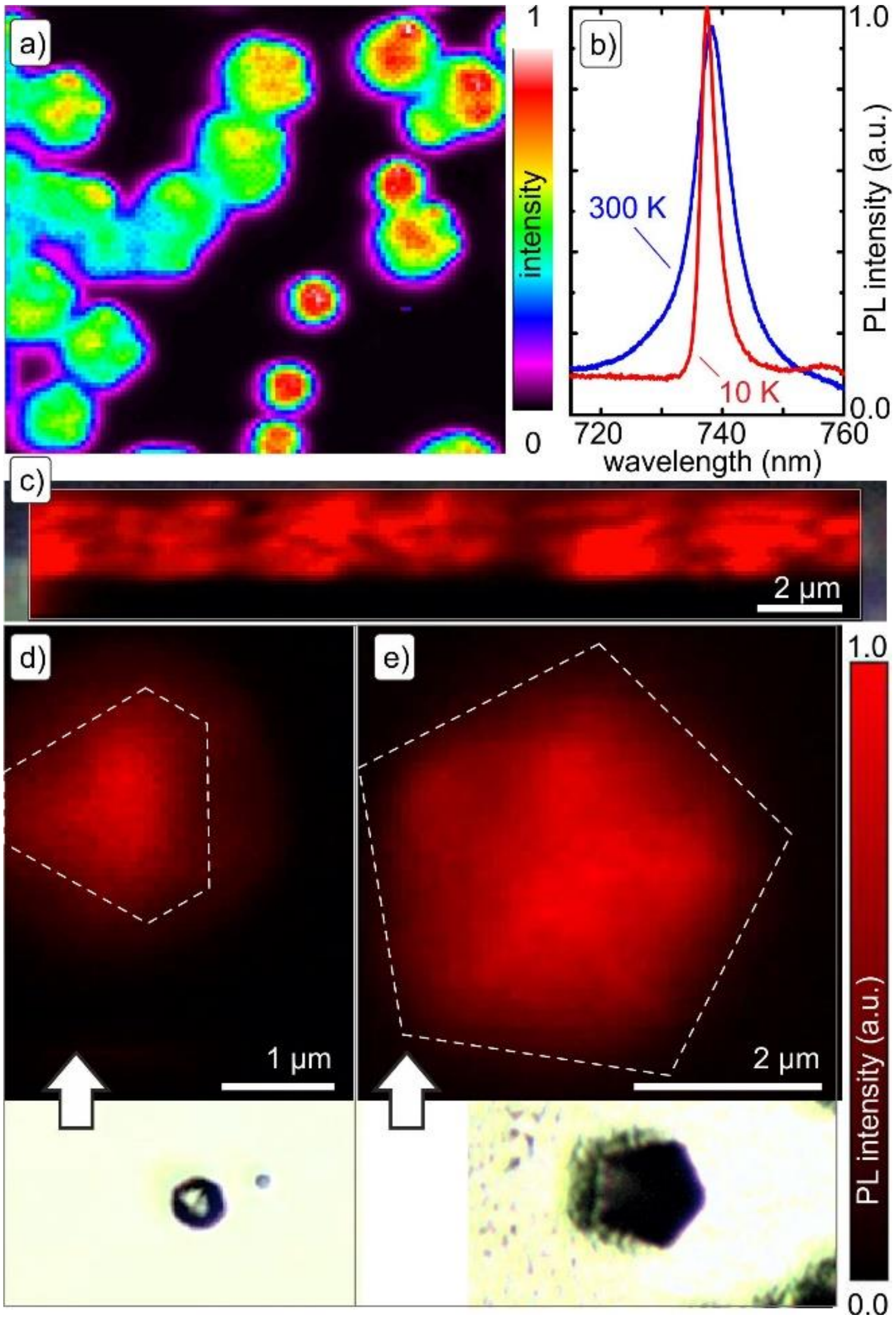

**FIG. 7.** µPL studies of diamond grains: a) overview µPL intensity map measured at 738 nm wavelength, b) typical PL spectra obtained on the isolated Ih-structures at 300 K and at 10 K with FWHM(SiV) of 4.8 nm and of 2.8 nm, respectively. c)-e) µPL intensity maps at 735 nm of c) MTP layer at a (110) cross-section, d) diamond Oh(111) pyramid, and e) 5-fold-twinned Dh-MTP measured at identical conditions. The insets show optical images of the investigated Oh- and Dh-structures (arbitrarily scaled).

ZPL emission segregate close to the MTP cores (see a cross-section µPL map, Fig. 7c). In order to get a deeper insight into the SiV spatial distribution within the MTP, high-resolution (HR-µPL) maps were recorded in "high-confocal" conditions allowing an in-plane resolution of ≈200 nm. Fig. 7d displays a µPL intensity map taken on top of a regular Oh(111) grain at λ=738 nm. The map shows a homogeneous distribution of the emission intensity with a characteristic tetragonal feature defined by a (111) core of the truncated pyramid. In contrast, the HR-µPL map of Dh-MTP (SEM image of similar structure is shown in Fig. 2a, Phase D) reveals a pronounced 5-fold symmetry (see Fig. 7e). The intense ZPL emission correlates perfectly with the positions of TBs in the Dh-MTP structure.

The HR-µPL observations are consistent with earlier studies (e.g. Ref. [24]) correlating extended defects and TBs in nanodiamond particles and layers with a steady narrowband of ZPL. The most probable explanation of the acquired data is a higher structural disorder in randomly bonded grain boundaries (e.g. presence of $sp^2$-C phase) altering the binding and position of Si atoms in relation to the vacancies in the diamond lattice. In contrast, the MTP-TBs provide a coherent (or partly coherent) interface for hosting impurity atoms, hence, resulting in the structural and energetic stability of SiV complexes.

Finally, to demonstrate an applicability of MTPs as micro-scale light sources, SiV-LEDs based on a vertical diamond/Schottky structure were manufactured using closed MTP layers, and investigated by µEL. This simple geometry was preferred over p-i-n [12] and planar Schottky [25] LED structures due to the simplicity in fabrication and sufficient vertical electrical conductivity of the MTP:Si films at 300 K (see Fig. 8a). The electrostimulated light is detected along the electrode's circumference (Fig. 8b). Here, the bright emitting spots correspond to the large MTP assemblies located near the edge of the metal layer. It fits well the observed surface morphology of the MTP-layer (see SEM image in Fig. 2b). On the other hand, optically active MTPs are discrete with a micrometer-scale distance between the intensively emitting spots. Therefore, we can conclude that not all the MTPs respond equally to the electrical stimulation. Electro-optical activity might relate to a charge carrier transport in the MTP-layer and this is a subject for future investigations.

The ZPL-emission from the SiV-LED was characterized by HR-µEL using a forward bias of $U_s = -200$ V ($I_s \approx -1.5$ mA) and an integration time of 100 ms. In Fig. 8c, the HR-µEL maps show the variation of the SiV-ZPL characteristics (intensity, FWHM of ZPL, and peak position) recorded over the area of a large MTP domain. According to the maps, the SiV centers located at TBs are the sources of the most intensive ZPL emission. Moreover, the peak position map shows that the SiVs located in the TB region have a higher energy of the ZPL transition indicating higher degree of the diamond lattice relaxation comparing to the core grain. These SiVs demonstrate also lower ZPL broadening with FWHM of 5-6 nm. Despite the fact that the whole MTP volume is excited in µEL contributing to the additional broadening of ZPL, this value is still comparable with those obtained from the "high-confocality" µPL measurements.

All aforementioned SiV emission properties originate from the specific TB features arising from twinning tetragonal crystal units in the MTPs. The corresponding crystallization mechanisms as well as their impact on the TBs' microstructure and location are discussed in the next section.

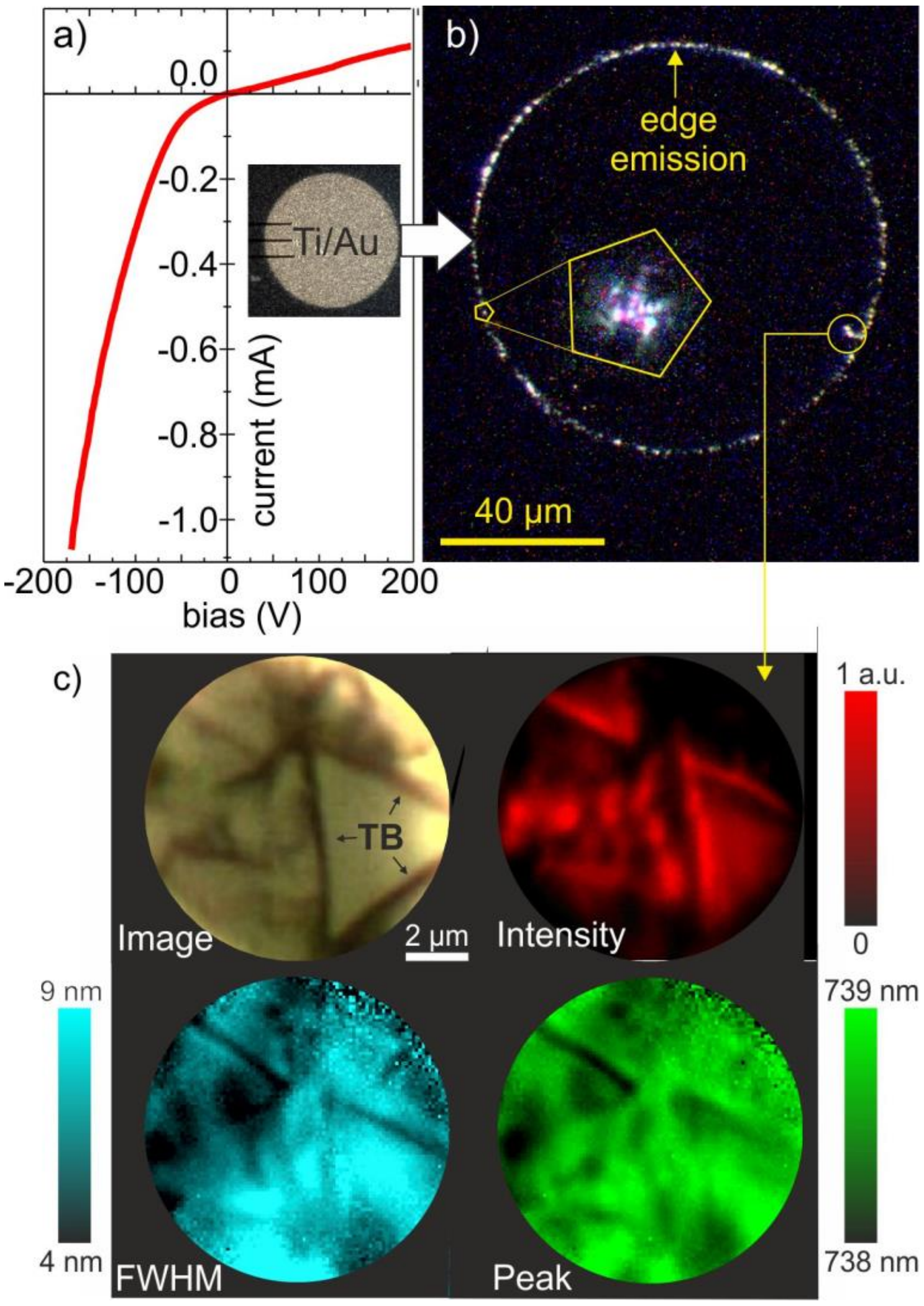

**FIG. 8.** µEL studies of the diamond MTP layer: a) voltage-current characteristic of Au/Ti/diamond Shottky diode; insets: optical images of the wire-bonded circular Au/Ti top-contact. b) Optical image of an intensive edge emission along the circumference of the circular electrode. c) Optical image and µEL maps of the ZPL emission originated from the TBs in closed MTP layer.

## IV. Discussion

### A. "Contact" and "penetration" twins

The terms "contact" and "penetration" twins are commonly used in CVD-diamond literature defining i) the twin/parent crystallographic relationship, and ii) the twin's interaction with the parent crystal [26]. In Ref. [13], we have discussed the contact twins inciting the Ih-MTP formation during the diamond crystallization on Ir(001). Such twins nucleate predominantly on re-entrant {111} facets by introducing a low-energy (111) stacking fault (SF) and continue their growth independently from the parent domain. Such behavior causes a series of secondary twin formations, and finally, completion of cyclic Dh- and Ih-MTPs containing 5 and 20 tetragonal segments, respectively. Correspondingly, all the {111}||{111}´

||{111}´´||… TBs containing single SF are nearly coherent having low density of misfit dislocations and dangling bonds.

In contrast to the growth on iridium, the MTP formation on silicon is governed by the penetration twins (PT, see section III.C and Fig. 4), which are only partly coherent with the parent crystal lattice. In Fig. 9a, an atomic arrangement in the area of the PT/parent interface is schematically shown using a (011)-plane projection (see Ref. [26] and references therein for more details).

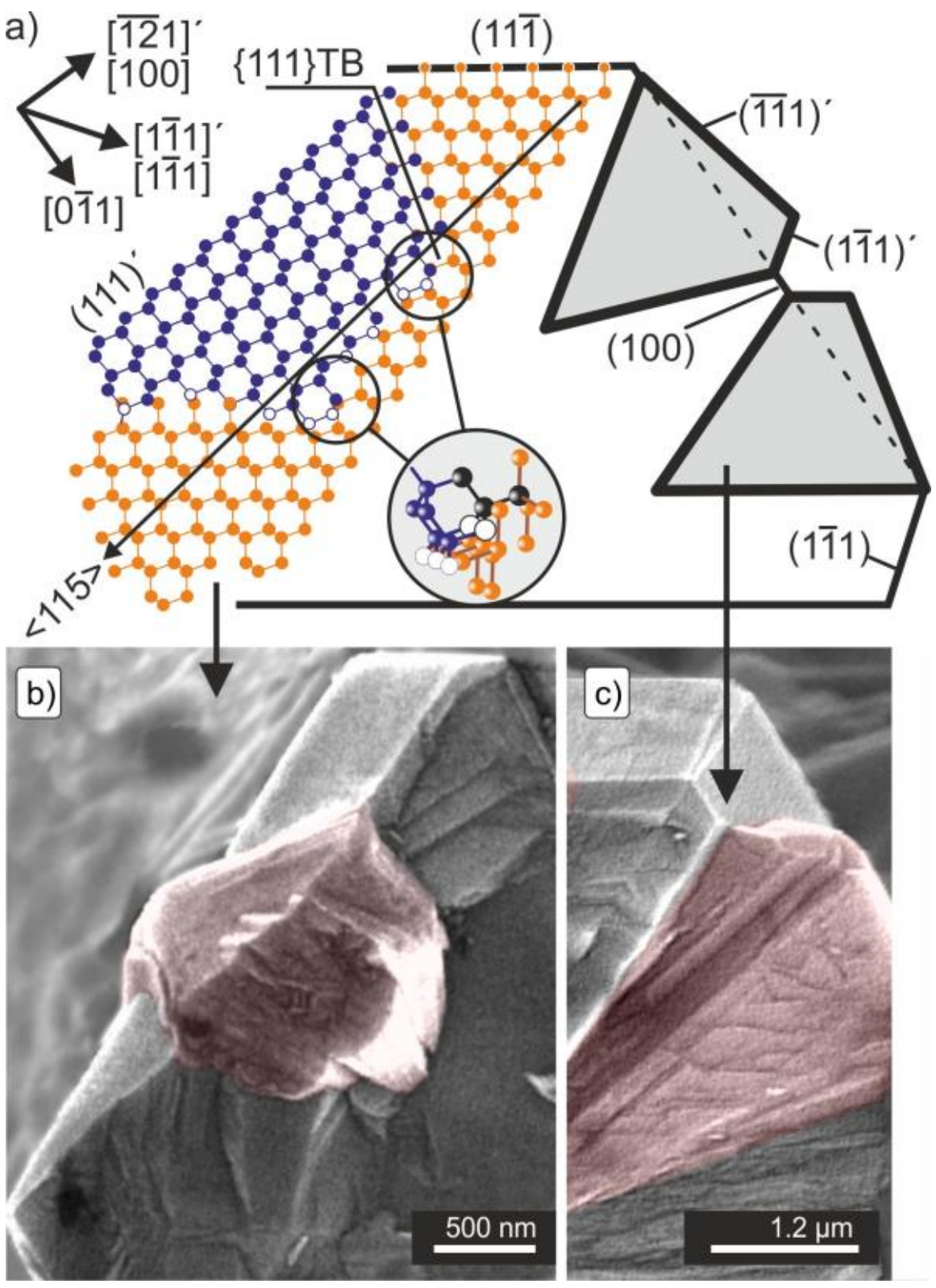

**FIG. 9.** (011) plane projection of the atomic arrangement in the area of PT nucleated on the parent {100} facet having atomic arrangement (111)´||(115). The PT and parent lattices shown by the blue and orange colors, respectively. Inset: 3D view of the twin-parent bonding zone. b)-c) Cross-sectional SEM micrographs of PT a) in the initial growth stage and b) in the later stage characteristic by well-defined TBs.

As demonstrated in Figs. 4 and 9b, at the initial PT growth phase, due to the *fcc* dangling bond configuration [27], four equivalent, orthogonal orientations of the PTs are possible. Moreover, for each of these four orientations, there is only single (111)´-plane, which can be coherently bound to the parent crystal via (111) SF. Remaining {111} TBs are partly incoherent (see inset in Fig. 9a), and such lattice incompatibility leads to high densities of the dangling bonds at the interfaces greatly increasing the interface energy ($E_{int}$).

Due to their "penetration" character, the PTs continue the growth competing with the parent crystal structure and develop from the nucleus (Fig. 9b) to the large domains having a well-defined TB structure (Fig. 9c). As revealed by SEM observations, in the growth stage preceding the MTP formation, a typical PT might retain a single coherent {111} TB (Fig. 9c), while the orientations of other TBs are undefined.

Hence, during their growth, the PTs generate excessively large area of incoherent TBs, which are capable to incorporate a significant amount of impurity atoms simultaneously retaining a high density of unsaturated chemical bonds (i.e. vacancies). It leads to vital consequences. In contrast to the MTPs grown on iridium in an impurity-free environment, the re-deposition of silicon atoms sputtered by the plasma from the substrate changes the growth dynamic and MTP properties on silicon prominently.

### B. Si as an anti-surfactant

There are two important aspects of MTP growth on Si by plasma CVD. The first one is an impact of Si impurities on the 2D mobility of the carbon species on diamond {100} surface. It is widely accepted that the diamond growth in $H_2$-plasma is a surfactant-mediated process. The atomic hydrogen saturates the carbon dangling bonds stabilizing $sp^3$-bonded structure on the surface due to the energy minimization [28]. Furthermore, Si adlayers were also reported to stabilize the $sp^3$-C structure during a 2D layer-by-layer growth in a hydrogen-free, ultra-high vacuum environment [29,30].

Plasma-CVD of diamond comprises a variety of highly competitive surface processes [16,31] complicating the understanding of the surface diffusion phenomena. However, taking into account the SIMS mapping results (see Fig. 6) revealing Si both in the bulk and in the boundaries, we cannot call Si a surfactant. Moreover, a direct comparison of the PT nucleation on Si- and Ir-substrates suggests that the Si impurities restrict tremendously the surface migration of adsorbed carbon species demonstrating a pure anti-surfactant behexperimentalavior. It fits early hypotheses claiming Si as an origin for multiply twin formation in CVD diamond-on-silicon layers [32].

Based on the presented SEM data, we can estimate a "capture" efficiency of the Si surface impurities, which, according to our model, kinetically control the PT nucleation immobilizing the adsorbed hydrocarbon radicals ($CH_3$, $CH_2$, CH, etc.) on {100} facets. In our consideration, $<\iota_{PT}>$ represents a characteristic length, which the $C_xH_y$ species travel until the chemisorption onto potential PT nucleation sites (i.e. impurity, vacancy complex, step bunching, edge dislocation, etc.). $<\iota_{PT}>$ is measured as a mean distance from the geometrical center of the largest twin-free {100} facet to its nearest edge. As discussed in section III.B (Fig. 2), $<\iota_{PT}>_{Si}$ of $C_xH_y$ radicals on Si-contaminated ({100}:Si) facets does not surpass 150 nm at T = 900 °C. For larger {100}:Si facets, the nucleation of at least one PT is registered. In contrast, the largest {100} twin-free area of the grains on Ir, crystallized at the identical thermodynamic conditions, does exceed 8×8 μm$^2$ resulting in $<\iota_{PT}>_C \approx 4$ μm [13]. Using the general expression for a thermally activated surface processes, the specific diffusion length $D_{PT} = (<\iota_{PT}>^2\nu)\exp(-E_a/kT)$, we can evaluate Si efficiency in a localization of $sp^3$-C on the PT nucleation sites by calculating the relative changes of the activation barrier on clean {100} and {100}:Si facets ($E_C$ and $E_{Si}$ , respectively). Here, ν is the atomic vibration frequency factor (≈$10^{13}$ $s^{-1}$ for monoatomic solids), and k is the Boltzmann constant. Assuming that the pre-exponential term ($<\iota_{PT}>^2\nu$) does not change its value upon a silicon addition, the observed reduction of $<\iota_{PT}>_{Si}$ should relate to the increase in a barrier height ($E_{Si}$-$E_C$) caused by the Si surface impurities. It accounts for an enormous increase in the barrier energy of ≈44 % driving the Si-assisted PT nucleation.

**C. SiVs segregation on diamond lattice irregularities**

The second aspect of the MTP growth on silicon relates to the segregation of Si atoms onto diamond lattice irregularities (i.e. extended defects and TBs), which leads to the PT nucleation and SiV formation, respectively. Si segregation on extended defects is often reported for nanodiamond [24], [33] as well as for other heteroepitaxial thin films, e.g. III-nitrides [15]. According to theoretical calculations [34], it is energetically unfavorable for the Si atom to occupy a substitutional position in the ideal diamond lattice. Moreover, in the interstitial state, Si may localize at the center of a double vacancy position [35] (see inset in Fig. 10), which is hardly realizable in the perfect diamond lattice due to a very large formation energy for a neutral vacancy site (≈7.1 eV for four broken bonds). In contrast, partly incoherent TBs provide the nucleation sites with a high-density of intrinsically created broken bonds (potential vacancies), and are capable to accommodate a high amount of Si atoms in a symmetric V-Si-V configuration.

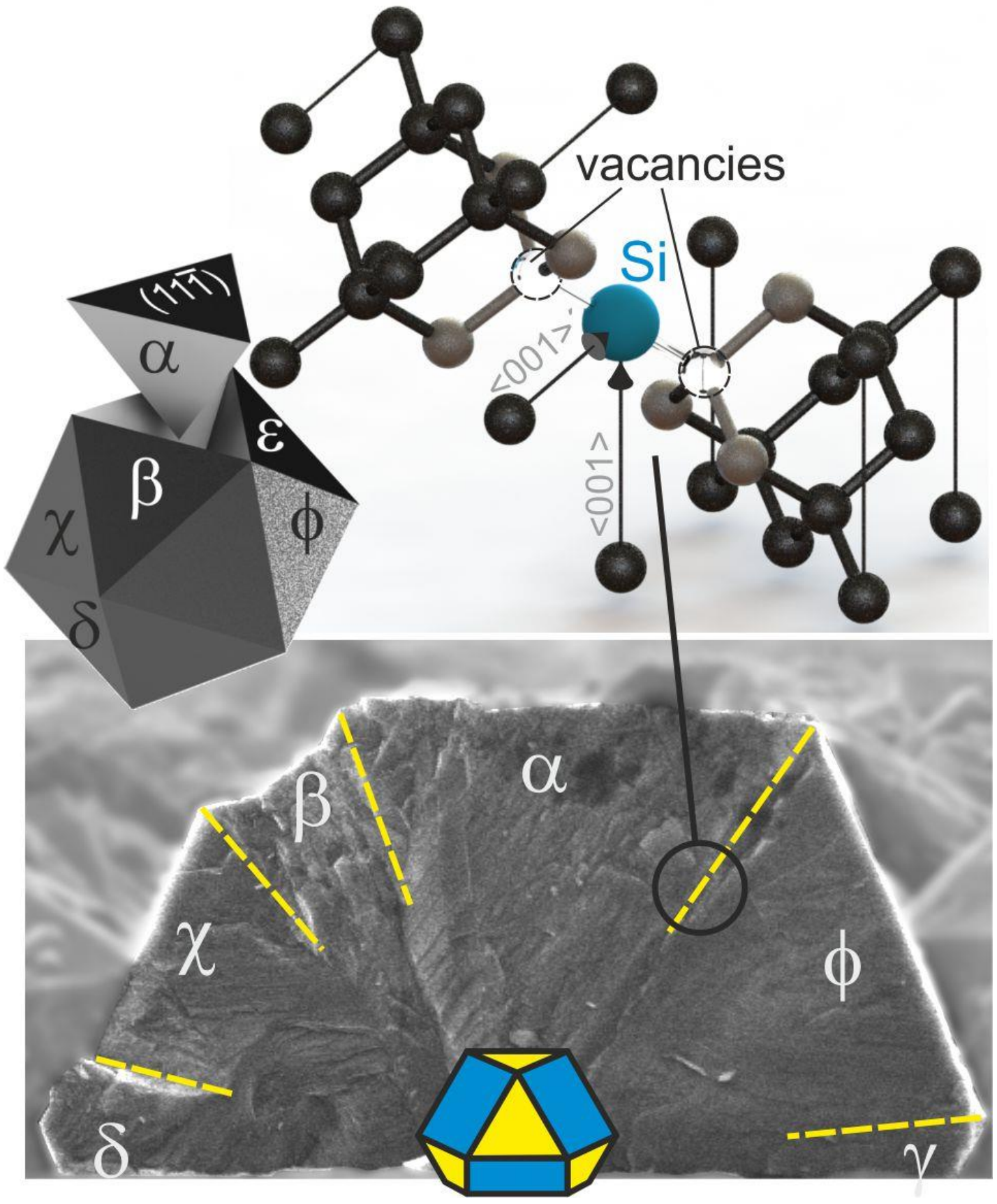

**FIG. 10.** a) SEM micrograph of the Ih-MTP close to the (011)-plane projection. The cross section shows Ih-segments arrangement and their crystallographic relationship to the parent crystal. Insets: schematic views of 20 tetragonal units forming Ih-structure (left), and {111} TB representing a possible atomic order at the (111) SF containing V-Si-V center (right).

Due to the uniaxial symmetry and the interstitial localization [36], the SiV centers are relatively insensitive to electric field fluctuations and phonon-mediated losses [37]. Moreover, at the SF area of {111} TB, the diamond lattice adjoining the SiV complexes have more degree of freedom for a short-range stress relaxation. It stabilizes the ZPL at the energy state close to the emission from the relaxed crystal. As a result, light emitted by SiV centers located within the TBs are characteristic by high intensity and narrow linewidth as demonstrated in the experimental section.

The MTP crystallization occurring after the PT formation stage is a thermodynamically driven process (see Ref. [13] for the discussion on the thermodynamic Wulff constructions). The Ih-cyclic structures develop multiply {111} TBs during their growth, which separate 20 tetragonal segments composing the MTP (Fig. 10). These TBs penetrate the whole MTP volume terminating at the Ih-ridges. As demonstrated by the experiments, the contribution from the SiVs localized at these ridges dominates in the ZPL emission preserving steady-state spectral features and long-time photo-stability of Si-doped MTPs – properties of the highest importance for the micro-scale monochromatic light sources.

## V. Final remarks and conclusions

Multiply twinned particles are very intriguing crystallographic objects with a number of unique properties originating from their symmetry and cyclic structure. MTPs are rarely applied in the modern optoelectronic devices, but their application potential is enormous owning to the “self-assembling” growth mode and variability of physical properties via in situ control of the crystallization. Especially, it concerns diamond MTPs hosting particular color centers – they might be a perfect solution for emerging ultra-compact optoelectronic and quantum optic applications.

Our aim in this work was to discuss specific properties of the diamond MTPs aggregating high-density of the silicon-vacancy complexes in the twin boundaries during growth. We demonstrated that the silicon impurities affect greatly the growth of Dh- and Ih-MTPs via intensive formation of the penetration twins on the {100} facets of octahedral grains. We also propose that by chemosorption of Si adatoms at the crystal irregularities, the growing MTPs minimize their interface energy stimulating extension of the {111} twin boundaries.

The most prominent feature of the twin boundaries is a steady-state atomic order defined by the crystallographic relationship of the twinned segments. It leads to the predictable positioning of the SiV complexes within the twin boundary in relation to the adjacent lattices, and results in the steady emission properties. The EL studies of SiV-LED structures based on closed MTP layers reveal exceptional photo-stability and constant spectral features of the ZPL emission over the whole SiV ensemble. These findings might be substantial for practical applications of diamond MTPs: the optically active SiVs are naturally located near the surface and their density and distribution in the boundaries can be controlled by the growth conditions.

## ACKNOWLEDGMENTS

The authors would like to thank M. Prescher, T. Fuchs and S. Klingelmeier for their technical assistance during the experiments. This research was partly financed from the European Union's Horizon 2020 research and innovation program under grant agreement No. 820374 (MetaboliQs).

**TABLE I.** Summary of the experimental conditions for carbonization, BEN and CVD processes.

| *Parameter* | *Carb.* | *BEN* | *CVD* |
|---|---|---|---|
| $T_{sub}$ (°C) | 900 | 680 | 900 |
| $W_{RF}$ (W) | 350 | 700 | 1400 |
| *P (Torr)* | 20 | 25 | 40 |
| $C_M$ (%) | 2.0 | 5.0 | 5.0 |
| $U_{DC}$ (V) | - | 250 | - |
| *t (min)* | 30 | 3 | 60-240 |

**TABLE II.** Five main crystallization phases occurring before the coalescence stage.

| | *Growth Phase Description* |
|---|---|
| *A* | growth of (111)-oriented truncated octahedrons (Oh) |
| *B* | spontaneous nucleation of the twins on clean {100} facets of the Oh(111) pyramids. |
| *C* | formation of 5-segment decahedron (Dh) structures sharing a common top {111} facet with the Oh(111) grains. |
| *D* | completion of Dh structure formation |
| *E* | completion and further growth of 20-segment Ihs. |

## Figure Captions

**FIG. 1.** SEM micrographs (75° view angle) of a) "rough" carbonized (111) Si surface and b) hetero-interface between diamond grains and silicon substrate taken after the cleavage along a <110> direction at the central area of the wafer.

**FIG. 2.** Top-view SEM micrographs demonstrating a) step-by-step transformation of Oh(111) pyramid (phase A) via "penetration twin" nucleation on {001} facets (phases B-C) into Dh-MTP (phase D) and then into Ih-MTP (phase E). b) SEM image (70° view angle) of a 2.5 µm thick, fiber textured diamond MTP layer at the stage of completed coalescence.

**FIG. 3.** XRD 2Θ/Θ scans of diamond/$Si_xC_y$/Si(111) samples containing: [1] - isolated (111)-oriented Oh and Ih grains (Fig. 2a); [2] - Ih-grains at the stage of partly completed coalescence (Fig. 2b); and [3] – thick diamond film showing polycrystalline structure. Insets: schematic representation of the layer morphologies corresponding to the plotted curves [1]-[3].

**FIG. 4.** SEM micrographs (70° - view angle) illustrating penetration tween (PT) nucleation and development. a) Initial formation of four orthogonally oriented PTs on a {100} facet of the Oh(111) grain: PT types A and B with a regular 4-fold symmetry, and PT types C, D, E developing a 5-fold (111) twinned structure on the reentrant side-walls. b) Close-up view of a partly developed 5-fold twinned structure. c) Cross-section view of the diamond grain demonstrating two types of PTs: regular one of type A and a twin of type F - the special case of PT penetrating the {111} facet. d) Schematic representation of Ih-related twins with respect to the parent Oh(111) geometry.

**FIG. 5.** Transmission electron microscopy studies of Ih-MTP: a) micrograph of the Ih-MTP containing epitaxially oriented core ($\alpha$), and two adjacent Ih-twins ($\varepsilon$ and $\phi$); b) HR-TEM image of $\phi$-twin having a high-density of stacking faults (SFs); c)-d) selective area diffraction patterns (SAED) taken c) at twin boundary area and d) at middle area of $\phi$-twin.

**FIG. 6.** a) SEM micrograph (70° - view angle, (110)-cleavage) of the investigated diamond MTP-film. b)-c) SIMS measurements carried out on closed MTP layers. b) SIMS depth-profiles for $^{12}C$ and $^{29}Si$ masses. c) Normalized in-plane mass distribution maps ($50\times50$ $\mu m^2$): position A – measurement point at a distance of $\approx$1 µm from the heterointerface, position B – diamond/Si heterointerface area; the dashed lines are a guide for the eye indicating approximate positions of the grain boundaries.

**FIG. 7.** µPL studies of diamond grains: a) overview µPL intensity map measured at 738 nm wavelength, b) typical PL spectra obtained on the isolated Ih-structures at 300 K and at 10 K with FWHM(SiV) of 4.8 nm and of 2.8 nm, respectively. c)-e) µPL intensity maps at 735 nm of c) MTP layer at a (110) cross-section, d) diamond Oh(111) pyramid, and e) 5-fold-twinned Dh-MTP measured at identical conditions. The insets show optical images of the investigated Oh- and Dh-structures (arbitrarily scaled).

**FIG. 8.** μEL studies of the diamond MTP layer: a) voltage-current characteristic of Au/Ti/diamond Shottky diode; insets: optical images of the wire-bonded circular Au/Ti top-contact. b) Optical image of an intensive edge emission along the circumference of the circular electrode. c) Optical image and μEL maps of the ZPL emission originated from the TBs in closed MTP layer.

**FIG. 9.** (011) plane projection of the atomic arrangement in the area of PT nucleated on the parent {100} facet having atomic arrangement (111)´||(115). The PT and parent lattices shown by the blue and orange colors, respectively. Inset: 3D view of the twin-parent bonding zone. b)-c) Cross-sectional SEM micrographs of PT a) in the initial growth stage and b) in the later stage characteristic by well-defined TBs.

**FIG. 10.** a) SEM micrograph of the Ih-MTP close to the (011)-plane projection. The cross section shows Ih-segments arrangement and their crystallographic relationship to the parent crystal. Insets: schematic views of 20 tetragonal units forming Ih-structure (left), and {111} TB representing a possible atomic order at the (111) SF containing V-Si-V center (right).